\newcommand{\CLASSINPUTinnersidemargin}{1 in} 
\documentclass[12pt,journal]{IEEEtran} 
\ifCLASSINFOpdf
\else
\fi

\usepackage{color}
\usepackage{amssymb}
\usepackage{amsthm}
\usepackage[cmex10]{amsmath}
\DeclareMathOperator{\E}{\mathbb{E}}

\newcommand {\Define} {\stackrel {\Delta} {=}  }

\newcommand{\mya}{\mathrel{\overset{\makebox[0pt]{{\tiny(a)}}}{=}}}

\newcommand {\pu} {p_{\text{u}}}

\newcommand {\lp} {l^{\prime}}

\usepackage{bm}
\usepackage{mathtools}
\usepackage{fixltx2e}
\usepackage{epsfig,makeidx,color}
\usepackage{graphicx,dblfloatfix}
\usepackage{amsbsy}
\usepackage{amssymb}
\usepackage{euscript}
\usepackage{lipsum}
\usepackage{cite}
\usepackage{chngcntr}
\usepackage{lineno}
\usepackage[T1]{fontenc}
\usepackage{microtype}
\usepackage{wasysym}
\usepackage{footnote}
\newtheoremstyle{dotless}{}{}{\itshape}{}{\bfseries}{:}{ }{}
\theoremstyle{dotless}

\newtheorem{lemma}{Lemma}

\newtheorem{remark}{\it Remark}

\usepackage[english]{babel}
    \usepackage[font=small,labelsep=space]{caption}
\usepackage{subcaption}    
\renewcommand{\baselinestretch}{1.8}
\makeatletter
\def\citenoauxwrite#1{\begingroup
\@fileswfalse
\cite{#1}\relax
\endgroup}
\makeatother

\begin{document}

\title{Impact of Frequency Selectivity on the Information Rate Performance of CFO Impaired Single-Carrier Massive MU-MIMO Uplink}
%
%
%

\author{Sudarshan~Mukherjee and 
        ~Saif Khan~Mohammed 
\thanks{Sudarshan Mukherjee and Saif Khan Mohammed are with the Department of Electrical Engineering, Indian Institute of Technology (I.I.T.) Delhi, India. Saif Khan Mohammed is also associated with Bharti School of Telecommunication Technology and Management (BSTTM), I.I.T. Delhi. Email: saifkmohammed@gmail.com. This work is supported by EMR funding from the Science and Engineering
Research Board (SERB), Department of Science and Technology (DST),
Government of India.}
}

\onecolumn
\maketitle

\vspace{-2.5 cm}

\begin{abstract}
In this paper, we study the impact of frequency-selectivity on the gap between the required per-user transmit power in the residual CFO scenario (i.e. after CFO estimation/compensation at the base-station (BS) from \protect\citenoauxwrite{gcom2015}) and that in the ideal/zero CFO scenario, for a fixed per-user information rate, in single-carrier massive MU-MIMO uplink systems with the TR-MRC receiver. Information theoretic analysis reveals that this gap decreases with increasing frequency-selectivity of the channel. Also, in the residual CFO scenario, an $\mathcal{O}(\sqrt{M})$ array gain is still achievable ($M$ is the number of BS antennas) in frequency-selective channels with imperfect channel estimates.
\end{abstract}

\vspace{-0.8 cm}
\begin{IEEEkeywords}

\vspace{-0.4 cm}
Massive MIMO, carrier frequency offset (CFO), frequency selective channel, Time-reversal maximum ratio combining (TR-MRC), single-carrier.
\end{IEEEkeywords}

\vspace{-0.8 cm}

%


\section{Introduction}
%
%
%
%

Massive multi-user (MU) multiple-input multiple-output (MIMO) system/large scale antenna system (LSAS) has been identified as one of the key next generation wireless technologies, due to its characteristic ability to provide huge increase in energy and spectral efficiency with increasing number of base-station (BS) antennas \cite{Andrews}. In a massive MU-MIMO system, the LSAS present in the BS serves an unconventionally large number of single-antenna user terminals (UTs) in the same time-frequency resource \cite{Marzetta1}. It has been shown that for a given number of UTs in a \textit{coherent} massive MU-MIMO system, with imperfect channel estimates, the required per-user transmit power (to achieve fixed per-user information rates) can be reduced as $\frac{1}{\sqrt{M}}$ with increasing $M$ (i.e. $\mathcal{O}(\sqrt{M})$ array gain\footnote[1]{Under the average power constraint, for a fixed $M$, fixed number of UTs and a fixed desired per-user information rate, the per-user average transmit power decreases with increasing $M$ \cite{Tse}.}), where $M$ is the number of BS antennas \cite{Ngo1}.

\par The result discussed above assumes perfect frequency synchronization between the BS and the UTs for coherent multi-user communication. In practice, acquisition of carrier frequency offsets (CFOs) between the carrier frequency of signals received from different UTs and the BS oscillator is a challenging task in massive MIMO systems (because of unconventionally large number of UTs). It has been observed that the existing optimal/near-optimal CFO estimation techniques for small scale MIMO systems are not amenable to practical implementation in massive MIMO systems due to prohibitive increase in their complexity with increasing number of UTs and also with increasing number of BS antennas \cite{Larsson2, gcom2015}. In \cite{Larsson2}, the authors consider an approximation to the joint maximum likelihood (ML) estimator for CFO estimation in frequency-flat massive MU-MIMO systems and analyze the information rate performance. However the CFO estimator presented in \cite{Larsson2} requires multi-dimensional grid search and hence has high complexity with increasing number of UTs. Also the information rate analysis  in \cite{Larsson2} does not consider the imperfect CSI scenario. Recently in \cite{gcom2015}, a low-complexity near-optimal CFO estimation technique has been proposed for massive MU-MIMO systems. Using this technique in \cite{tvt2016}, the information rate performance in frequency-flat Rayleigh fading channel has also been studied. However for frequency-selective single-carrier\footnote[2]{It has been shown that a single-carrier massive MU-MIMO system can achieve performance comparable to their OFDM counterparts \cite{Pitarokoilis}.} massive MU-MIMO uplink systems, it is not known as to how the information theoretic performance gap between the ideal/zero CFO scenario and the residual CFO scenario (i.e. with CFO estimation/compensation from \cite{gcom2015}) would vary with increasing frequency-selectivity. In this paper, we study this performance gap and show the interesting result that it decreases with increasing frequency-selectivity.

\par The novel contributions of our work presented in this paper are as follows: (i) we have derived a closed-form expression for per-user information rates of each UT with the TR-MRC\footnote[3]{TR-MRC (Time-reversal maximum ratio combining) receiver is a well-known \textit{low-complexity single-carrier multi-user detector in frequency-selective}  massive MIMO channels \cite{Phasenoise}.} receiver at the BS in the imperfect CSI scenario, in the presence of residual CFO errors (after CFO compensation using the CFO estimator in \cite{gcom2015}); (ii) analysis of the information rate expression reveals that an $\mathcal{O}(\sqrt{M})$ array gain is still achievable in \textit{frequency-selective channel} with CFO estimation/compensation, i.e., no loss in array gain when compared to the ideal/zero CFO scenario; (iii) further analysis of the information rate expression reveals the interesting result that for a fixed per-user information rate, the gap in the required per-user transmit power between the residual CFO scenario and the ideal/zero CFO scenario \emph{decreases with increasing frequency-selectivity} (i.e. number of channel memory taps $L$) of the channel. For instance with $M = 160$ BS antennas, $K = 10$ UTs and a per-user rate of $3$ bits per channel use (bpcu), this gap in per-user transmit power is approximately $4.22$ dB when $L = 1$ and is approximately $0.07$ dB when $L = 20$. [\textbf{{Notations:}} $\E$ denotes the expectation operator. $(.)^{\ast}$ denotes the complex conjugate operator.]

\begin{figure}[t]
\vspace{-0.5 cm}
\centering
\includegraphics[width= 5 in, height= 1.3 in]{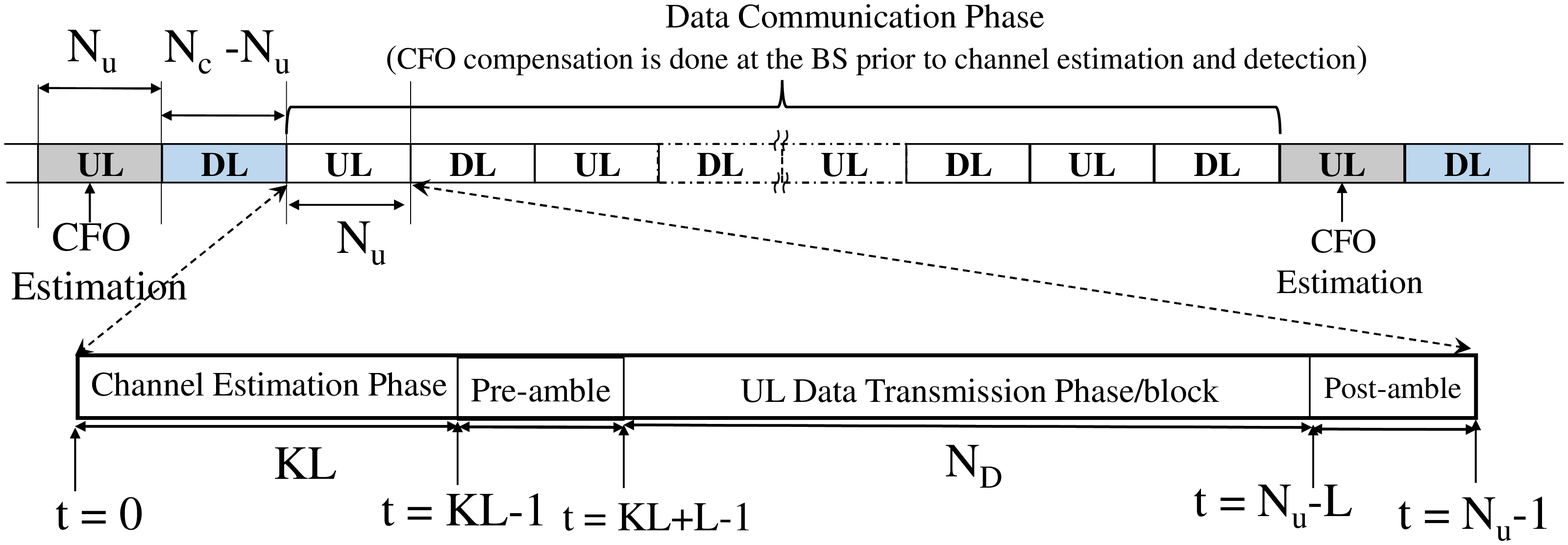}
\caption {The communication strategy: CFO Estimation and Compensation Strategies and Data Communication.} 
\label{fig:commstrat}
\vspace{-0.8 cm}
\end{figure}

\vspace{-0.8 cm}

\section{System Model \& CFO Estimation}
We consider a single-cell single-carrier massive MU-MIMO BS with $M$ antennas serving $K$ single antenna UTs in the same time-frequency resource. The baseband frequency selective channel is modelled as a discrete time finite impulse response (FIR) filter with $L$ channel taps. The channel gain coefficient from the $k^{\text{th}}$ UT to the $m^{\text{th}}$ BS antenna at the $l^{\text{th}}$ channel tap is given by $h_{mk}[l] \, \sim \mathcal{C}\mathcal{N}(0, \sigma_{hkl}^2)$, where $\{\sigma_{hkl} > 0\}$ models the power delay profile (PDP) of the channel for the $k^{\text{th}}$ UT ($k = 1, 2, \ldots, K$; $l = 0, 1, \ldots, L-1$ and $m = 1, 2, \ldots, M$). Since a massive MU-MIMO system is expected to operate in time-division duplexed (TDD) mode, each coherence interval ($N_c$ channel uses) is split into an uplink (UL) slot ($N_u$ channel uses) and a downlink (DL) slot ($N_c - N_u$ channel uses). From the coherent communication strategy depicted in Fig.~\ref{fig:commstrat}, it is observed that CFO estimation is performed in a special UL slot prior to the conventional UL communication (channel estimation/UL data transmission). 


\par For the CFO estimation phase, the UTs transmit pilots having average power $\pu$ for $N \leq N_u$ channel uses. The pilot sequence is assumed to be divided into $B \Define \lceil N/KL \rceil$ blocks of $KL$ channel uses each. The $k^{\text{th}}$ UT transmits an impulse of amplitude $\sqrt{KL \pu}$ at time $t = (b -1)KL + (k - 1)L +l$ and zero elsewhere ($b = 1, 2, \ldots, B$; $l = 0,1, \ldots, L-1$ and $k = 1, 2, \ldots,K$). Using block-wise correlation of the received pilots, CFO estimation is performed at the BS, using the low-complexity CFO estimator proposed in \cite{gcom2015}. Let $\omega_k$ be the CFO of the $k^{\text{th}}$ UT and $\widehat{\omega}_k$ be the estimate of $\omega_k$. It can be shown from the central limit theorem (CLT), that as $M \to \infty$, the error in CFO estimation is asymptotically Gaussian, i.e., $(\widehat{\omega}_k - \omega_k) \sim \mathcal{N}(0, \sigma_{\omega_k}^2)$. Here $\sigma_{\omega_k}^2 \Define \E[(\widehat{\omega}_k - \omega_k)^2]$ is the mean squared error (MSE) of CFO estimation for the $k^{\text{th}}$ UT and is given by \cite{gcom2015}

\vspace{-0.8 cm}

\begin{IEEEeqnarray}{rCl}
\label{eq:msecfo}
\sigma_{\omega_k}^2 & = & \frac{\frac{1}{\gamma_k}\Big(\frac{G_k}{B - 1} + \frac{1}{2K \gamma_k}\Big)}{M(N - KL)(KL)^2G_k^2} \, ,
\IEEEeqnarraynumspace
\end{IEEEeqnarray}

\vspace{-0.3 cm}

\noindent where\footnote[4]{Note that the low-complexity CFO estimator in \cite{gcom2015} is well defined if and only if $|\omega_k KL| \ll \pi$. For most practical massive MIMO systems, this condition would hold true \cite{gcom2015}.} $\gamma_k \Define \frac{\pu}{\sigma^2}\sum_{l = 0}^{L - 1}\sigma_{hkl}^2$ is the received SNR from the $k^{\text{th}}$ ($\sigma^2$ is the AWGN power at the BS) and $G_k \Define {\sum_{m = 1}^{M}\sum_{l = 0}^{L - 1}|h_{mk}[l]|^2}/{(M \sum_{l = 0}^{L-1}\sigma_{hkl}^2)}$. Note that with $M \to \infty$, from the strong law of large numbers, for independent $h_{mk}[l]$, $\lim\limits_{M \to \infty} G_k = 1$.


\begin{remark}
\label{mseM}
(Proposition~2 from \cite{gcom2015})
\normalfont From the expression of $\sigma_{\omega_k}^2$ in \eqref{eq:msecfo} it is clear that with increasing $M \to \infty$, fixed $N$, $K$ and $L$, the required received SNR $\gamma_k$ (to achieve a fixed desired MSE of CFO estimation) decreases as $\frac{1}{\sqrt{M}}$, i.e., with $\gamma_k \propto \frac{1}{\sqrt{M}}$, we have $\lim\limits_{M \to \infty}\sigma_{\omega_k}^2 = \text{constant}$. \hfill \qed
\end{remark}

\vspace{-0.8 cm}

\section{Uplink Data Communication}
\vspace{-0.1 cm}

After the CFO estimation phase, the conventional UL data communication starts at $t = 0$ (see Fig.~\ref{fig:commstrat}). Note that the CFO compensation is performed at the BS prior to UL channel estimation and also prior to UL receiver processing. For UL data communication, the first $KL$ channel uses are dedicated for UL pilot transmission, which is then followed by ($L-1$) channel uses of pre-amble sequence. The UL data transmission occurs in the next $N_D$ channel uses and is followed by $(L-1)$ channel uses of post-amble transmission.\footnote[5]{The symbols transmitted in the pre- and post-amble sequences are independent and identically distributed (i.i.d.) and are assumed to have the same distribution as the information symbols (see section III-B) to ensure the correctness of the achievable information rate expression.}

\vspace{-0.6 cm}

\subsection{Channel Estimation}

In the channel estimation phase, the UTs transmit sequentially in time, i.e., the $k^{\text{th}}$ UT transmits an impulse of amplitude $\sqrt{KL \pu}$ at time $t = (k - 1)L$ and zero elsewhere. The received pilot at the $m^{\text{th}}$ BS antenna at $t = (k-1)L +l$ is given by $r_m[(k-1)L + l] = \sqrt{KL \pu}\, h_{mk}[l]e^{j\omega_k[(k-1)L + l]} + w_m[(k-1)L+l]$, where $m = 1, 2, \ldots, M$; $l = 0,1, \ldots, L-1$ and $k = 1, 2, \ldots, K$. Here $w_m[(k-1)L + l] \sim \mathcal{C}\mathcal{N}(0, \sigma^2)$ is the circular symmetric AWGN. We first perform CFO compensation for the $k^{\text{th}}$ UT by multiplying $r_m[(k-1)L +l]$ with $e^{-j\widehat{\omega}_k [(k-1)L+l]}$, which is then followed by computing the maximum likelihood estimate of the channel gain coefficient, i.e., $\widehat{h}_{mk}[l] \Define \frac{1}{\sqrt{KL \pu}}r_m[(k-1)L +l]e^{-j\widehat{\omega}_k[(k-1)L+l]} = \widetilde{h}_{mk}[l] + \frac{1}{\sqrt{KL \pu}} n_{mk}[(k-1)L + l]$, where\footnote[6]{{Both $h_{mk}[l]$ and $w_m[(k-1)L+l]$ have uniform phase distribution (i.e. circular symmetric) and are independent of each other. Clearly, rotating these random variables by fixed angles (for a given realization of CFOs and its estimates) would not affect the distribution of their phases and they will remain independent. Therefore the distribution of $\widetilde{h}_{mk}[l]$ and $n_{mk}[(k-1)L+l]$ would be same as that of $h_{mk}[l]$ and $w_m[(k-1)L+l]$ respectively.}} $\widetilde{h}_{mk}[l] \Define h_{mk}[l] e^{-j \Delta \omega_k[(k-1)L +l]} \sim \mathcal{C}\mathcal{N}(0, \sigma_{hkl}^2)$ is the effective channel gain coefficient and $n_{mk}[(k-1)L + l] \Define w_m[(k-1)L + l]e^{-j\widehat{\omega}_k[(k-1)L+l]} \sim \mathcal{C}\mathcal{N}(0, \sigma^2)$. Here $\Delta \omega_k \Define \widehat{\omega}_k - \omega_k$ is the residual CFO error.

\vspace{-0.6 cm}

\begin{figure*}[t]
\normalsize
\vspace*{-25pt}
\begin{IEEEeqnarray}{lCl}
\label{eq:isi}
\text{ISI}_k[t]  =  \sqrt{\pu}\sum\limits_{m=1}^{M} \sum\limits_{l = 0}^{L-1} \sum\limits_{\lp = 0, \lp \neq l}^{L-1} \, \widetilde{h}_{mk}^{\ast}[l]\, \widetilde{h}_{mk}[\lp]\, x_k[t-\lp+l]\, e^{-j\Delta \omega_k (t - (k-1)L)}\, e^{-j\Delta \omega_k (l - \lp)}\\
\label{eq:mui}
\text{MUI}_k[t]  =  \sqrt{\pu}\sum\limits_{m=1}^{M} \sum\limits_{l = 0}^{L-1} \sum\limits_{q=1, q \neq k}^{K} \sum\limits_{\lp = 0}^{L-1} \, \widetilde{h}_{mk}^{\ast}[l]\, \widetilde{h}_{mq}[\lp]\, x_q[t-\lp+l]\, e^{j((\widehat{\omega}_q - \widehat{\omega}_k)(t+l)-\Delta \omega_q (t - (q-1)L+(l - \lp)))}\, \\
\IEEEeqnarraynumspace
\label{eq:corrsignoise}
\nonumber \hspace{-0.6 cm} \E\Big[\text{ES}_k [t]W_k^{\ast}[t]\Big] = \E[A_k[t]] \Bigg[ \underbrace{\E\Big[x_k[t]\Big\{A_k^{\ast}[t]x_k^{\ast}[t] - \E[A_k[t]]x_k^{\ast}[t]\Big\}\Big]}_{= \, 0,\, \text{since}\,\, x_k[t]\, \text{and}\, A_k[t] \, \text{are independent}} \,+\, \underbrace{\E\bigg[x_k[t] \Big\{\text{ISI}_k [t] + \text{MUI}_k [t]\Big\}^{\ast}\bigg]}_{= \,\,0,\,\, \text{since all}\,\, x_k[t]\,\, \text{are i.i.d.}}\,\\
\,\, \hspace{10 cm}+ \underbrace{\E\bigg[x_k[t]\text{EN}_k^{\ast}[t]\bigg]}_{\substack{=\,\, 0 \,\, (\text{since} \,\, \text{all} \, n_{mk}[t] \, \text{are i.i.d.,}\\ \text{ zero mean and independent}\\ \text{of }x_k[t])}}\Bigg] \, = \, 0.
\IEEEeqnarraynumspace
\end{IEEEeqnarray}
\hrulefill
\vspace*{-15pt}
\end{figure*}

\subsection{Uplink Receiver Processing}

After channel estimation and preamble transmission, the UL data transmission begins at time $t = KL+L-1$. Let $x_k[t] \sim\mathcal{C}\mathcal{N}(0,1)$ be the i.i.d. information symbol transmitted by the $k$-th UT at the $t$-th channel use and $\pu$ be the average per-user transmit power. The received signal at the $m$-th BS antenna at time $t$ is therefore given by $r_m[t] = \sqrt{\pu} \sum_{q=1}^{K} \sum_{l=0}^{L-1}h_{mq}[l] x_q[t-l]e^{j\omega_qt} + w_m[t]$, where $t = KL+L-1, \ldots, KL+L+N_D-2 (= N_u-L)$. To detect $x_k[t]$, we first perform CFO compensation for the $k^{\text{th}}$ UT on the received signal, followed by TR-MRC processing \cite{Phasenoise}. Output of the TR-MRC receiver at time $t$ is given by

\vspace{-1.1 cm}

\begin{IEEEeqnarray}{rCl}
\label{eq:detsig}
\widehat{x}_k[t] & \Define & \sum\limits_{m=1}^{M}\sum\limits_{l=0}^{L-1} \widehat{h}_{mk}^{\ast}[l] \underbrace{ r_m[t+l] \, e^{-j\widehat{\omega}_k(t+l)}}_{\text{CFO Compensation}} \mya A_k[t]\, x_k[t]\, + \text{ISI}_k[t] + \text{MUI}_k[t] + \text{EN}_k[t],
\IEEEeqnarraynumspace
\end{IEEEeqnarray}

\vspace{-0.4 cm}

\noindent where step $(a)$ follows from the expressions of $\widehat{h}_{mk}[l]$ (see Section III-A) and $r_m[t]$ (see Section III-B) and $A_k[t] \Define \sqrt{\pu} \, \sum\limits_{m=1}^{M} \sum\limits_{l =0}^{L-1} |\widetilde{h}_{mk}[l]|^2 \, e^{-j\Delta \omega_k (t - (k-1)L)}$. Here the terms $\text{ISI}_k[t]$ (inter-symbol interference), $\text{MUI}_k[t]$ (multi-user interference) are given by \eqref{eq:isi}-\eqref{eq:mui} at the top of the last page and $\text{EN}_k[t] \Define \widehat{x}_k[t] - A_k[t]x_k[t] - \text{ISI}_k[t] - \text{MUI}_k[t]$ ($\widehat{x}_k[t]$ is given by the first line of \eqref{eq:detsig}). From \eqref{eq:detsig}, we therefore have

\vspace{-1.5 cm}

\begin{IEEEeqnarray}{rCl}
\label{eq:xkhat}
\widehat{x}_k[t] & = & \underbrace{\E[A_k[t]] \, x_k[t]}_{\Define \, \text{ES}_k[t]} + \underbrace{\text{SIF}_k[t] + \text{ISI}_k[t] + \text{MUI}_k[t] + \text{EN}_k[t]}_{\Define \, W_k[t]},
\IEEEeqnarraynumspace
\end{IEEEeqnarray}

\vspace{-0.4 cm}

\noindent where $\text{SIF}_k[t] \Define (A_k[t] - \E[A_k[t]])x_k[t]$ is the time-varying self-interference component and $\text{ES}_k[t]$ is the effective signal component.\footnote[7]{In \eqref{eq:xkhat}, $\E[.]$ is taken across multiple channel realizations and also across multiple CFO estimation phases.} Note that the statistics of both $\text{ES}_k[t]$ and $W_k[t]$ are functions of $t$. However for a given $t$, the realization of $W_k[t]$ is i.i.d. across multiple UL data transmission blocks (i.e. coherence intervals). Therefore for each $t$, we have a SISO (single-input single-output) channel in \eqref{eq:xkhat}, when viewed across multiple coherence intervals. Thus for $N_D$ channel uses, we have $N_D$ SISO channels with distinct channel statistics. We therefore have $N_D$ different channel codes, one for each of these $N_D$ channels. The data received in the $t^{\text{th}}$ channel use of every coherence interval is jointly decoded at the BS.\footnote[8]{This coding strategy has also been used in \cite{Phasenoise,tvt2016}.}

\begin{savenotes}
\begin{table}[b]
\caption[position=top]{{\textsc{List of Variance of all components of $\text{W}_k[t]$.}}}
\label{table:vartable}
\centering
\begin{tabular}{| c | c |}
\hline
Component & Variance\\ 
\hline
\vspace{-0.3 cm} & \\
$\text{ES}_k[t]$ & $M^2\pu \Big(\sum\limits_{l=0}^{L-1} \sigma_{hkl}^2\Big)^2 \, e^{-\sigma_{\omega_k}^2(t - (k-1)L)^2}$\\
\hline
\vspace{-0.3 cm} & \\
$\text{SIF}_k[t]$ & $\substack{M^2 \pu \Big(\sum\limits_{l=0}^{L-1} \sigma_{hkl}^2\Big)^2 \, \Big[1 - e^{-\sigma_{\omega_k}^2(t - (k-1)L)^2}\Big]\\+M \pu \sum\limits_{l=0}^{L-1} \sigma_{hkl}^4}$ \\
\hline
\vspace{-0.3 cm} & \\
$\text{ISI}_k[t]$ & $M \pu \Big[\Big(\sum\limits_{l=0}^{L-1} \sigma_{hkl}^2\Big)^2 - \sum\limits_{l=0}^{L-1} \sigma_{hkl}^4\Big]$\\
\hline
\vspace{-0.3 cm} & \\
$\text{MUI}_k[t]$ & $M \pu \Big(\sum\limits_{l=0}^{L-1} \sigma_{hkl}^2\Big)\Big(\sum\limits_{q = 1, q \neq k}^{K} \sum\limits_{l^{\prime} = 0}^{L-1} \sigma_{hq\lp}^2 \Big)$\\
\hline
\vspace{-0.3 cm} & \\
$\text{EN}_k[t]$ & $\frac{M\sigma^2}{K}\left(\sum\limits_{q=1}^{K}\sum\limits_{l=0}^{L-1}\sigma_{hql}^2\right) + \frac{M\sigma^4}{Kp_{\text{u}}}+ M\sigma^2 \left(\sum\limits_{l=0}^{L-1}\sigma_{hkl}^2\right)$\\
\hline
\end{tabular}
\end{table}
\end{savenotes}

\vspace{-0.5 cm}
%
%
\subsection{Achievable Information Rate}

Since $x_k[t]$ and $n_{mk}[t]$ are all independent and zero mean, it can be shown that $\text{ES}_k[t]$, $\text{SIF}_k[t]$, $\text{ISI}_k[t]$, $\text{MUI}_k[t]$ and $\text{EN}_k[t]$ are all zero mean and uncorrelated with one another (see \eqref{eq:corrsignoise} at the top of the page). Since the overall noise and interference term $W_k[t]$ and $\text{ES}_k[t]$ are uncorrelated, a lower bound on the information rate for the effective channel in \eqref{eq:xkhat} can be obtained by considering the worst case uncorrelated additive noise (in terms of mutual information). With Gaussian information symbols $x_k[t]$, this worst case uncorrelated noise is also Gaussian with the variance $\E[|\text{W}_k[t]|^2]=\E[|\text{SIF}_k[t]|^2]+ \E[|\text{ISI}_k[t]|^2] +\E[|\text{MUI}_k[t]|^2] + \E[|\text{EN}_k[t]|^2]$ \cite{Hasibi2}. Therefore an achievable lower bound on the information rate for the $t^{\text{th}}$ channel is given by $I(\widehat{x}_k[t]; x_k[t]) \geq \log_2(1 + \text{SINR}_k[t])$, where $\text{SINR}_k[t] \Define \E[|\text{ES}_k[t]|^2]/\E[|W_k[t]|^2]$. Therefore the information rate for the $k^{\text{th}}$ UT is given by $I_k = \frac{1}{N_u} \sum_{t = KL+L-1}^{N_u-L} \log_2(1 + \text{SINR}_k[t])$. Using the expressions of $\text{ES}_k[t]$, $\text{SIF}_k[t]$, $\text{ISI}_k[t]$, $\text{MUI}_k[t]$ and $\text{EN}_k[t]$ from \eqref{eq:isi},\eqref{eq:mui}, \eqref{eq:detsig} and \eqref{eq:xkhat}, and $\Delta \omega_k = (\widehat{\omega}_k - \omega_k) \sim \mathcal{N}(0, \sigma_{\omega_k}^2)$, the variances of each term are computed and summarized in Table~\ref{table:vartable}. Using Table~\ref{table:vartable}, $\text{SINR}_k[t] = \E[|\text{ES}_k[t]|^2]/\E[|\text{W}_k[t]|^2]$ is given by

\vspace{-1.2 cm}

\begin{IEEEeqnarray}{rCl}
\label{eq:sinr}
\text{SINR}_k[t] & = & \frac{e^{-\sigma_{\omega_k}^2(t - (k-1)L)^2}}{[1 - e^{-\sigma_{\omega_k}^2(t - (k-1)L)^2}] + \frac{1}{MK\gamma_k^2} + \frac{c_1}{M \gamma_k} + \frac{c_2}{M}},
\IEEEeqnarraynumspace
\end{IEEEeqnarray}

\vspace{-0.3 cm}

\noindent where $c_1 \Define 1 + \frac{\sum_{q=1}^{K}\theta_q}{K\theta_k}$ and $c_2 \Define \frac{1}{\theta_k}\sum_{q=1}^{K} \theta_q$. Here $\theta_k \Define \sum_{l=0}^{L-1} \sigma_{hkl}^2$. 

\begin{remark}
\label{arraygain}(Array Gain)
\normalfont In the following, for a fixed desired information rate, and therefore fixed $\text{SINR}_k[t]$, we examine the rate of decrease in the required $\gamma_k$ with increasing $M$. On the RHS of \eqref{eq:sinr} we note that the numerator and the first term in the denominator depend on the received SNR $\gamma_k$ only through the MSE of CFO estimation, $\sigma_{\omega_k}^2$. From Remark~\ref{mseM} we know that if $\gamma_k \propto \frac{1}{\sqrt{M}}$, then as $M \to \infty$, $\sigma_{\omega_k}^2$ converges to a constant, i.e., the numerator and the first term in the denominator of \eqref{eq:sinr} converge to constant values. Further as $M \to \infty$, the last term in the denominator of \eqref{eq:sinr}, i.e, $\frac{c_2}{M}$ vanishes. The rest two terms, $\big(\frac{1}{MK\gamma_k^2} + \frac{c_1}{M \gamma_k}\big)$ however depend on both $M$ and $\gamma_k$. Since with decreasing $\gamma_k$, the term $\frac{1}{MK\gamma_k^2}$ would eventually dominate the other term $\frac{c_1}{M \gamma_k}$, we must therefore decrease $\gamma_k$ as $\frac{1}{\sqrt{M}}$ so that the $\text{SINR}_k[t]$ converges to a constant as $M \to \infty$ (This shows that with every doubling in $M$, $\gamma_k$ decreases roughly by $1.5$ dB for a fixed per-user rate when $M \to \infty$ (see the change in $\gamma_k$ from $M = 320$ to $M = 640$ in Table~\ref{table:varM}).\hfill \qed
\end{remark}

\begin{savenotes}
\begin{table}[b]
\caption[position=top]{{{Minimum required $\gamma_k$ (in dB) for fixed information rate $I_k = 1$ bpcu ($k = 1$) with increasing $M$, $K = 10$ and $L = 10$.\footnote[9]{The system parameters for data in Table~\ref{table:varM} is given in the first paragraph of Section IV.}}}}
\label{table:varM}
\centering
\begin{tabular}{| c | c | c | c| c|}
\hline
$M = 40$ & $M = 80$ & $M = 160$ & $M = 320$ & $M = 640$\\ 
\hline
\vspace{-0.65 cm} & \\
-9.5266 & -12.2927 & -14.5049 & -16.4462 & -18.2341 \\
\hline
\end{tabular}
\end{table}
\end{savenotes}

%


\vspace{-0.2 cm}

\begin{lemma}
\label{varL}
\normalfont Consider $|\omega_k KL| \ll \pi$ and $\lim\limits_{M \to \infty} M \gamma_k^2 = \text{constant}>0$. With fixed $K$, $N$ and a fixed desired information rate of the $t^{\text{th}}$ channel code $R_k[t] \Define \lim\limits_{M \to \infty} \log_2(1 + \text{SINR}_k[t]) \gg R_{0,k}[t]$ \big(where $R_{0,k}[t] \Define \log_2(1 + \alpha_{k,t})$, $\alpha_{k,t} \Define \frac{(t - (k-1)L)^2}{2(N - KL)(KL)^2}$ \big), the asymptotic (i.e. $M \to \infty$) gap between the required $\gamma_k$ in the residual CFO scenario (i.e. after CFO estimation/compensation) and that in the ideal/zero CFO scenario ($\gamma_{k,0}$), i.e., $\lim\limits_{M \to \infty}\frac{\gamma_k}{\gamma_{k,0}}$ decreases with increasing $L$, provided $L \leq \frac{N}{2K}$.
\end{lemma}

\begin{IEEEproof}
Since $\lim\limits_{M \to \infty}M \gamma_k^2 =  \text{constant}>0$ and $\lim\limits_{M \to \infty}G_k = 1$, from \eqref{eq:msecfo} we have $\lim\limits_{M \to \infty}\sigma_{\omega_k}^2(t - (k-1)L)^2 = \frac{\lim\limits_{M \to \infty}\frac{1}{MK\gamma_k^2}(t - (k-1)L)^2}{2(N-KL)(KL)^2} = \alpha_{k,t} \theta_k$, where $\theta_k \Define\lim\limits_{M \to \infty}\frac{1}{MK\gamma_k^2}$. Using this limit in \eqref{eq:sinr}, we have $R_k[t] = \lim\limits_{M \to \infty}\log_2(1+\text{SINR}_k[t]) = \log_2(1 + e^{-\alpha_{k,t}\theta_k}/(1 - e^{\alpha_{k,t}\theta_k} + \theta_k))$. Let $\theta_k = \theta^{\prime}$ be the unique solution to this equation for a fixed $R_k[t]$, i.e.,

\vspace{-1.4 cm}

\begin{IEEEeqnarray}{rCl}
\label{eq:eqn1}
(1 + \theta_k)(1 - 2^{-R_k[t]}) = e^{-\alpha_{k,t} \theta_k}.
\IEEEeqnarraynumspace
\end{IEEEeqnarray}

\vspace{-0.5 cm}

\noindent Therefore $(1 + \theta^{\prime})(1 - 2^{-R_k[t]}) = e^{-\alpha_{k,t} \theta^{\prime}} < 1 \implies \theta^{\prime} < \frac{1}{2^{R_k[t]} - 1}\ll \frac{1}{2^{R_{0,k}[t]} - 1} = \frac{1}{\alpha_{k,t}} \implies \alpha_{k,t} \theta^{\prime} \ll 1 \implies e^{-\alpha_{k,t}\theta_k} \approx 1 - \alpha_{k,t} \theta_k$. Substituting this in \eqref{eq:eqn1} with $\theta_k = \theta^{\prime}$, we have $\frac{1}{\theta^{\prime}} = \alpha_{k,t} + (1 + \alpha_{k,t})(2^{R_k[t] - 1}) \approx (1 + \alpha_{k,t})(2^{R_k[t] - 1})$ ($\because R_k[t] \gg R_{0,k}[t] > \log_2(1 + \alpha_{k,t}/(1 + \alpha_{k,t}))$). Similarly for the zero CFO scenario, using $\sigma_{\omega_k}^2 = 0$ in \eqref{eq:sinr} we have $\theta_0 \Define \lim\limits_{M \to \infty} \frac{1}{MK\gamma_{k,0}^2} = \frac{1}{2^{R_k[t]}-1}$. From the expressions of $\theta_0$ and $\theta^{\prime}$, the asymptotic SNR gap is given by 

\vspace{-1 cm}

\begin{IEEEeqnarray}{rCl}
\label{eq:snrgap}
\lim\limits_{M \to \infty}\frac{\gamma_k}{\gamma_{k,0}} = \lim\limits_{M \to \infty} \sqrt{\frac{1/MK\gamma_{k,0}^2}{1/MK\gamma_k^2}} = \sqrt{\frac{\theta_0}{\theta^{\prime}}} = \sqrt{1 + \alpha_{k,t}}.
\IEEEeqnarraynumspace
\end{IEEEeqnarray}

\vspace{-0.2 cm}

\indent Since $\frac{N}{KL} \geq 2$ \cite{gcom2015}, it follows that $\alpha_{k,t} = \frac{(t - (k-1)L)^2}{2(N - KL)(KL)^2}$ monotonically decreases with increasing $L \leq \frac{N}{2K}$. Hence from \eqref{eq:snrgap} it follows that $\lim\limits_{M \to \infty}\frac{\gamma_k}{\gamma_{k,0}}$ decreases with increasing $L$. \hfill \IEEEQEDhere

\end{IEEEproof}

Lemma~\ref{varL} shows the interesting result that the SNR gap between the residual CFO scenario and the zero CFO scenario decreases with increasing frequency-selectivity ($L$) of the channel. For the residual CFO scenario, with $\gamma_k = \frac{c}{\sqrt{M}}$, we note that $\text{SINR}_k[t]$ depends only on $c$ and $L$, for sufficiently large $M$ ($\because \frac{c_1}{M\gamma_k}$ and $\frac{c_2}{M}$ in \eqref{eq:sinr} vanish with $M \to \infty$). From \eqref{eq:msecfo} it is clear that for fixed $c$, the MSE decreases with increasing $L$ and hence $\text{SINR}_k[t]$ would increase. Similarly from \eqref{eq:sinr} we also note that for a fixed $L$ and decreasing $c$, the MSE increases and therefore $\text{SINR}_k[t]$ decreases. Hence, for a fixed desired information rate, i.e. fixed $\text{SINR}_k[t]$, we must decrease $c$ with increasing $L$, i.e., $\gamma_k = \frac{c}{\sqrt{M}}$ must be decreased. Therefore the SNR gap with the zero CFO scenario decreases with increasing $L$ (since with $\sigma_{\omega_k}^2 = 0$, it is clear from \eqref{eq:sinr} that $\text{SINR}_k[t]$ is independent of $L$). This conclusion is also supported in Fig.~\ref{fig:infoL}.

\vspace{-0.4 cm}

\section{Numerical Results and Discussions}

In this section, through Monte-Carlo simulations, we study the variation in the minimum required received SNR $\gamma_k$ with increasing per-user information rate for different values of $L$ (fixed $N$, $K$, $M$ and $N_u$). We assume the following: carrier frequency $f_c = 2$ GHz, a maximum CFO of $\kappa f_c$ ($\kappa = 0.1$ PPM) and communication bandwidth $B_{\text{w}} = 1$ MHz. Thus $|\omega_k| \leq 2\pi \kappa \frac{f_c}{B_{\text{w}}} = \frac{\pi}{2500}$. At the start of every CFO estimation phase, the CFOs $\omega_k$ ($k = 1, 2, \ldots, K$) assume new values (independent of the previous values) uniformly distributed in $[-\frac{\pi}{2500},\frac{\pi}{2500}]$. The duration of uplink is $N_u = 2000$ channel uses and pilot length for CFO estimation $N = N_u$. The PDP is the same for all UTs and is given by $\sigma_{hkl}^2 = 1/L$, $l = 0, 1, \ldots, L-1$; $k = 1, 2, \ldots, K$. The information rate is computed using \eqref{eq:sinr}, with $\sigma_{\omega_k}^2 = \E[(\widehat{\omega}_k - \omega_k)^2]$ replaced by its expression in \eqref{eq:msecfo} with $G_k = 1$ (see the line before Remark~\ref{mseM}).

\begin{figure}[!t]
\vspace{-0.3 in}
\centering
\includegraphics[width= 4.5 in, height= 2.1 in]{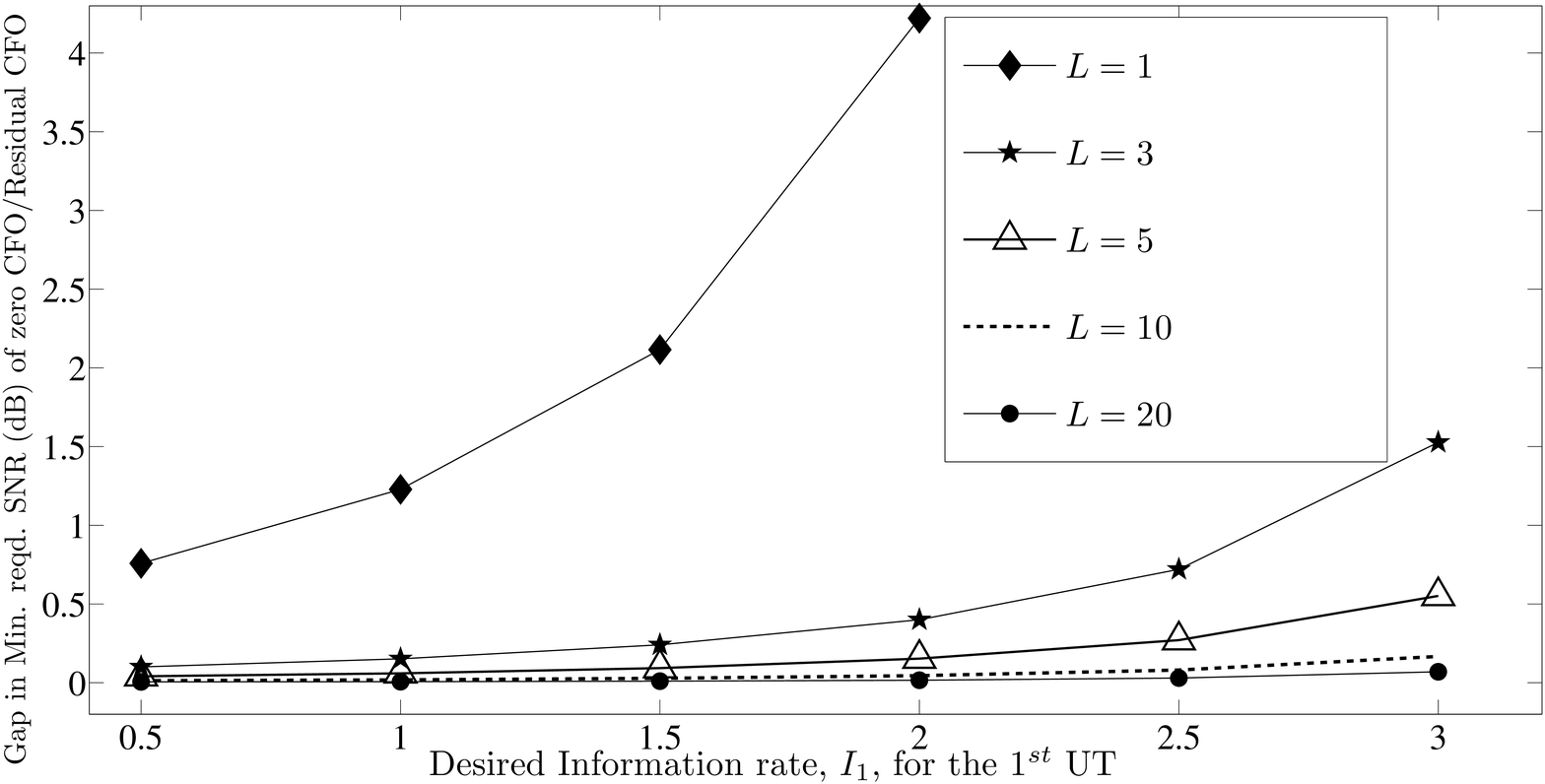} 
\caption {Plot of the SNR gap between the residual CFO and ideal/zero CFO scenarios with increasing information rate, $I_k$ for the first user ($k = 1$), for different $L = 1, 3, 5, 10$ and $20$. Fixed parameters: $M = 160$, $K = 10$ and $N = N_u = 2000$.}
\label{fig:infoL}
\vspace{-0.9 cm}
\end{figure}

\par In Fig.~\ref{fig:infoL} we depict the variation in the SNR gap between the residual CFO scenario and the ideal/zero CFO scenario, with increasing desired information rate $I_k$ for the first UT ($k = 1$) for $M = 160$ BS antennas and $L = 1, 3, 5, 10$ and $20$. Note that with $L = 1$ for $I_k = 3$ bpcu, the SNR gap is $\approx 4.22$ dB, which quickly decreases to a small value of $\approx 0.07$ dB when $L = 20$. This supports our conclusion in Lemma~\ref{varL} that the performance of the TR-MRC receiver in massive MIMO systems with CFO estimation/compensation proposed in \cite{gcom2015}, approaches the zero CFO scenario performance limit with increasing frequency-selectivity of the channel.


%

%


\ifCLASSOPTIONcaptionsoff
  \newpage
\fi



%

\vspace{-0.7 cm}
\bibliographystyle{IEEEtran}
\bibliography{IEEEabrvn,mybibn}

\begin{thebibliography}{10}
\providecommand{\url}[1]{#1}
\csname url@samestyle\endcsname
\providecommand{\newblock}{\relax}
\providecommand{\bibinfo}[2]{#2}
\providecommand{\BIBentrySTDinterwordspacing}{\spaceskip=0pt\relax}
\providecommand{\BIBentryALTinterwordstretchfactor}{4}
\providecommand{\BIBentryALTinterwordspacing}{\spaceskip=\fontdimen2\font plus
\BIBentryALTinterwordstretchfactor\fontdimen3\font minus
  \fontdimen4\font\relax}
\providecommand{\BIBforeignlanguage}[2]{{%
\expandafter\ifx\csname l@#1\endcsname\relax
\typeout{** WARNING: IEEEtran.bst: No hyphenation pattern has been}%
\typeout{** loaded for the language `#1'. Using the pattern for}%
\typeout{** the default language instead.}%
\else
\language=\csname l@#1\endcsname
\fi
#2}}
\providecommand{\BIBdecl}{\relax}
\BIBdecl

\bibitem{Andrews}
J.~Andrews, S.~Buzzi, W.~Choi, S.~Hanly, A.~Lozano, A.~Soong, and J.~Zhang,
  ``What {W}ill 5{G} {B}e?'' \emph{{IEEE} J. Sel. Areas Commun.}, vol.~32,
  no.~6, pp. 1065--1082, June 2014.

\bibitem{Marzetta1}
T.~Marzetta, ``Noncooperative {C}ellular {W}ireless with {U}nlimited {N}umbers
  of {B}ase {S}tation {A}ntennas,'' \emph{{IEEE} Trans. Wireless Commun.},
  vol.~9, no.~11, pp. 3590--3600, November 2010.

\bibitem{Tse}
D.~Tse and P.~Viswanath, \emph{Fundamentals of {W}ireless
  {C}ommunications}.\hskip 1em plus 0.5em minus 0.4em\relax Cambridge:
  Cambridge University Press, 2005.

\bibitem{Ngo1}
H.~Q. Ngo, E.~Larsson, and T.~Marzetta, ``Energy and {Spectral} {Efficiency} of
  {Very} {Large} {Multiuser} {MIMO} {Systems},'' \emph{{IEEE} Trans. Commun.},
  vol.~61, no.~4, pp. 1436--1449, April 2013.

\bibitem{Larsson2}
H.~Cheng and E.~Larsson, ``Some {F}undamental {L}imits on {F}requency
  {S}ynchronization in massive {MIMO},'' in \emph{Signals, Systems and
  Computers, 2013 Asilomar Conference on}, Nov 2013, pp. 1213--1217.

\bibitem{gcom2015}
S.~Mukherjee and S.~K. Mohammed, ``Low-{C}omplexity {CFO} {E}stimation for
  {M}ulti-{U}ser {M}assive {MIMO} {S}ystems,'' in \emph{Global Communications
  Conference (GLOBECOM), 2015 IEEE}, Dec 2015, pp. 1--7.

\bibitem{tvt2016}
S.~Mukherjee, S.~K. Mohammed, and I.~Bhushan, ``Impact of {CFO} {E}stimation on
  the {P}erformance of {ZF} {R}eceiver in {M}assive {MU-MIMO} {S}ystems,'' to
  appear in \textit{IEEE Transactions on Venhicular Technology}.

\bibitem{Pitarokoilis}
A.~Pitarokoilis, S.~K. Mohammed, and E.~G. Larsson, ``On the {O}ptimality of
  {S}ingle-{C}arrier {T}ransmission in {L}arge-{S}cale {A}ntenna {S}ystems,''
  \emph{{IEEE} Wireless Commun. Lett.}, vol.~1, no.~4, pp. 276--279, August
  2012.

\bibitem{Phasenoise}
A.~Pitarokoilis~et al, ``Uplink {P}erformance of {T}ime-{R}eversal {MRC} in
  {M}assive {MIMO} {S}ystems {S}ubject to {P}hase {N}oise,'' \emph{{IEEE}
  Trans. Wireless Commun.}, vol.~14, no.~2, pp. 711--723, Feb 2015.

\bibitem{Hasibi2}
B.~Hassibi and B.~Hochwald, ``How much training is needed in multiple-antenna
  wireless links?'' \emph{{IEEE} Trans. Inf. Theory}, vol.~49, no.~4, pp.
  951--963, April 2003.

\end{thebibliography}


\end{document}